\documentstyle[11pt,aaspp4]{article}

\lefthead{Wen, Cui, \& Bradt}
\righthead{X-ray intensity--hardness  correlation in Cyg~X-1} 

\begin{document}

\title{The Correlated Intensity and Spectral Evolution of Cyg X-1 During State Transitions}
 \author{Linqing Wen\altaffilmark{1}, Wei Cui\altaffilmark{1,2}, and Hale V. Bradt\altaffilmark{1}}

\altaffiltext{1}{Center for Space Research, MIT, Cambridge, MA 02139
 USA; lqw@space.mit.edu, cui@physics.purdue.edu, hale@space.mit.edu }
 \altaffiltext{2}{Department of Physics, Purdue University, West Lafayette, IN 47907 }

\begin{abstract}
Using data from the All-Sky Monitor aboard the {\em Rossi X-ray Timing
Explorer} ({\it RXTE}), we found that the 1.5--12 keV X-ray count rate  
of Cyg X-1 is, on time scales from 90 seconds to at least 10 days, 
strongly correlated with the spectral hardness of the source in the 
soft state, but is weakly anti-correlated with the latter in the hard 
state. The correlation shows an interesting evolution during the 1996 
spectral state transition. The entire episode can be roughly divided 
into three distinct phases: (1) a 20-d transition phase from the hard 
state to the soft state, during which the correlation changes from 
being negative to positive, (2) a 50-d soft state with a steady 
positive correlation, and (3) a 20-d transition back to the hard state. 
The pointed {\it RXTE} observations confirmed the ASM results but
revealed new behaviors of the source at energies beyond the ASM 
pass band. We discuss the implications of our findings. 

\end{abstract}


\section{Introduction}  
Cyg X-1 is identified with a binary system of $5.6$-day orbital period
which contains an O$9.7$ Iab supergiant and a compact object that is
believed to be a black hole (\cite{bolton72}; \cite{webster72}).
Observations  indicate that the system usually assumes
one of the two states, the hard state and the soft state (\cite{oda77}; 
\cite{nolan84}; \cite{tanaka95}).  Most of the time, Cyg X-1 stays in 
the hard state, where its soft X-ray (often $2$--$10$ keV) flux is 
relatively low and the X-ray spectrum is hard.  Every few years, the 
system undergoes a transition from the hard state to the soft
state,  during which the soft X-ray flux increases, often by a factor 
of more than $4$, and the X-ray spectrum softens. It remains in the 
soft state for weeks to months before returning to the hard state. 
There is a strong anti-correlation between the soft and the hard 
(e.g., $>$ 20 keV) X-ray  flux. Consequently, the bolometric X-ray 
luminosity does not vary significantly (\cite{zhang97a}). The transition 
between the two states lasts from less than a day to more than a 
week. 

A number of models have been proposed to explain the spectral
evolution of Cyg X-1 during the state transitions. For instance,
Ichimaru (1977) suggested that the physical condition of the accreted
gas near the disk outer-boundary could drive the disk into either an
optically thick state or an optically thin state, which correspond to
the soft and hard state respectively. Zhang, Cui, \& Chen (1997)
argued, based on the effects of black hole rotation,
that the state transition of Cyg X-1 may be caused by a temporary
reversal of the disk from being retrograde (hard state) to prograde
(soft state), which can occur in wind accretion systems (e.g.,
\cite{matsuda87}; \cite{ruffert97}). In the magnetic flare model 
(\cite{matteo99}), the
soft (hard) state corresponds to a lower (higher) scale height of
magnetic flares above the accretion disk. The flares are energized by
the reconnection of magnetic flux tubes rising from the accretion disk
due to magnetic buoyancy instability.  In the soft state, intense
flares close to the disk greatly enhance the soft photon field which
results in a soft X-ray spectrum. In the hard state, the flare is
triggered high above the disk, the system is ``photon starved'' and
thus results in a hard, Comptonized spectrum. In the framework of
advection-dominated accretion flows (ADAFs), Esin et al. (1998) argued
that the spectral states of Cyg X-1 are uniquely determined by the
mass accretion rate \.{\it m}. In the hard state, \.{\it m} is
relatively low, the inner edge of the thin disk is far away from the
black hole and thus the emission from the disk is weak compared to
that from the large, optically thin ADAF region. In the soft state,
the \.{\it m} is higher, which causes the ADAF region to shrink and
the thin disk to extend closer to the black hole. Therefore, at low
energies the emission from the disk dominates over that from the 
smaller ADAF region. More data are needed to distinguish these models.

In this Letter, we report the results from a quantitative study of the
correlation between the X-ray flux and the spectral properties of 
Cyg X-1, on  time scales of 90 s to at least 10 days, in both the hard 
and soft states, as well as during the transition phases. 

\section{Data}

The primary data set for this investigation comes from the All-Sky
Monitor (ASM) on board the {\em Rossi X-ray Timing Explorer} ({\it RXTE})
(\cite{bradt93}). The ASM yields the intensities of the sources
observed in units of the count
rate in three energy bands ($1.5$--$3$, $3$--$5$, and
$5$--$12$ keV).  The $1.5$--$12$ keV Crab Nebula
flux is about $75$ c/s.  Typically, a source is observed for a 90 s
exposure $\sim 15$ times a day,
which provides sufficient coverage for studying phenomena
such as the $\sim 90$ day state transition episode of Cyg X-1.    A 
detailed description of the
ASM and the light curves can be found in Levine et al. (1996) and
Levine (1998).

We used the ASM observations of Cyg X-1 between
1996 March and 2000 April,  which cover the entire  90-day 1996 soft
state including its 
transition  phases (Cui et al. 1997a; Zhang et al. 1997) and four
years of the hard state.   The count rate in
the 1.5--12 keV band and two hardness ratios HR1 and HR2 were
 computed.  
HR1 is the ratio of the count rate in the $3$--$5$ keV
band to that in the $1.5$--$3$ keV band, and HR2 is the ratio in the 
$5$--$12$ keV band to that in the $3$--$5$ keV band. The hardness 
ratios provide a rough measure of the X-ray spectral shape of the 
source.

The higher-quality data from the Proportional Counter Array (PCA)
(\cite{jahoda96}) aboard {\it RXTE} (with much poorer coverage) were
used to verify the ASM results and to study the flux-spectral hardness
evolution over a wider energy range.  We used 63 PCA observations of Cyg
X-1, with exposure times ranging from 700 s to 22 ks (but typically 
3 ks), obtained from 1996 March 26 to 1998 April 28, nineteen of which
cover the 1996 spectral state transition.  To minimize the known
effects of calibration uncertainties at higher energies, we limited
the spectral analyses to the energy range 2.5--25 keV. Also, since the
number of the proportional counter units (PCUs) that were turned on varied
from observation to observation, we used data only from PCU 0 which
was active during all of the selected observations, in order to
facilitate comparison of different observations.

For each PCA observation,  we constructed a light curve with 100 s 
time bins and an energy spectrum for each time bin using FTOOLS (v. 5.0).  
Each spectrum was then fit with an empirical model using XSPEC (v. 10.0). 
For the soft state, the model consists of a broken power-law component, 
an iron line around 6.4 keV and a fixed absorption 
($n_H =5.6 \times 10^{21}$ cm~$^{-2}$, e.g., \cite{ebisawa96}, 
\cite{cui97a}). For the hard state, we replaced the broken power law 
with a simple power law for the continuum, except for observations 
near the state transition or a pronounced soft flare in 1997 June. In
the latter cases, the broken power law, sometimes an additional 
blackbody component, is required for obtaining an adequate fit. Variable 
absorbing column density is also necessary for some of the observations 
near superior conjunction of the X-ray source. All fits have reduced 
$\chi ^2 < 1.5 $. Note that our main objective here is simply to derive 
the energy flux of the source and its hardness ratios, as opposed to 
finding  a physical model for the observed spectrum. The flux was calculated 
in the  2.5--25 keV band, as well as in the 3--5 keV, 5--12 keV, 12--17
keV, and 17--25 keV bands.  The energy bands were chosen such that the 
first two coincide roughly with the two upper ASM bands. Here, we 
defined the the hardness ratios as the ratios of energy fluxes between 
the first two (5--12 keV/3--5 keV) and the last two (17--25 keV/12--17 keV) 
energy bands.

\section{Analysis and Results}

The ASM light curve of Cyg X-1 clearly indicates long term variability
on time scales ranging from 90 s to hundreds of days (Fig.~\ref{corr}, 
see also Fig.~1 in Wen et al. 1999).  The rms variation of the light 
curve with a 90 s exposure in the 1.5--12 keV band is about 35\% in
the hard state, and 22\% in the soft state; both are much larger than
the expected uncertainties (7\% and 3\% respectively) in the data
based on counting statistics and systematic uncertainties.

We quantify the correlation between the source count rate and spectral
hardness by means of a non-parametric method first proposed by
Spearman (the Spearman ranking method; \cite{press92}). For each data
set, the value of each data point is replaced by the value of its
rank among all other data points.  That is, for N data points, the
smallest value would be replaced with value 1 and the largest with N.
If some of the data points have identical values, they were assigned
the mean of the ranks they would have if they were to be slightly
different. The correlation coefficient $r_s$ is defined as
\begin{equation}
r_s=\frac{\sum _i(R_i-\overline{R})(S_i-\overline{S})}{\sqrt{\sum
 _i(R_i-\overline{R})^2}\sqrt{\sum _i(S_i-\overline{S})^2}},  
\end{equation}
where $R_i$ and $S_i$ are the assigned ranks for the data points in
each of the two data sets. The significance of a non-zero value was
tested by computing:
\begin{equation}
t=r_s\sqrt{\frac{N-2}{1-r^2_s}}
\end{equation}
which is distributed approximately as Student's distribution with
$N-2$ degrees of freedom. 

The main advantage of the Spearman-ranking method is that the
significance level of the correlation does not depend on the the
original probability distribution of the data. Because of this, the
significance level of the correlation can be reliably computed even if
the number of the data points  is small. Despite some loss of
information in replacing the original values by ranks, the method is
reliable in the sense that when a correlation is demonstrated to be
present non-parametrically, it is really there. On the other hand, 
there exist examples
where correlations could  be detected parametrically but could not be
detected non-parametrically. However, such examples are believed to be
very rare in practice (\cite{press92}).  

We then computed, using eq.~1, the correlation coefficients between
the X-ray count rate  and the spectral hardness for the ASM and the PCA data.
For the ASM data with its original 90 s time bins, we calculated each
correlation coefficient over a ``correlation interval'' for several
values that ranged from 1 to 20 days.  To explore the correlation
relations for variability of different time scales, the calculations
were repeated for time bins up to 10 days for the soft state, and up
to 100 days for the hard state.  In these cases, the correlation
intervals were chosen to be 5 to 20 times the time bin size such that
the average number of data points per interval is greater than three.
For the PCA data (with 100 s time bins), we calculated one correlation
coefficient for each observation, due to the limited number of data 
points. In all our calculations, data segments with less than 3 data 
points were excluded.

\subsection{The ASM results} 
\begin{figure}[f]
\begin{centering}
\epsfxsize=5.0in \epsfbox{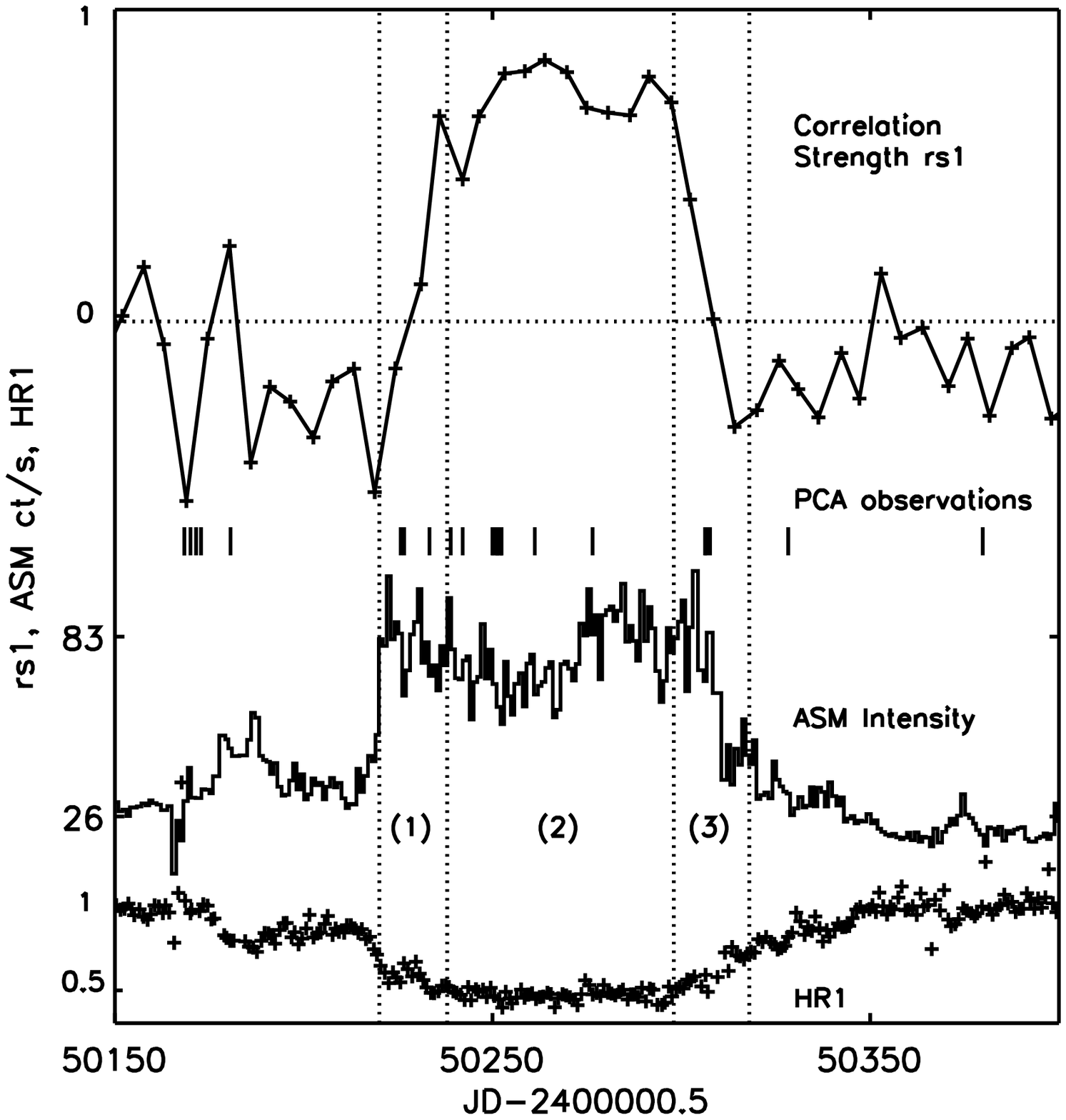}
\end{centering}
\figcaption[fig1.eps]{Correlation coefficients of the 1.5-12 keV ASM
 count rate and the hardness ratio HR1 for  90 s time bins and a 5.6 day
correlation interval (see text). Also shown are the ASM count rate and
hardness ratio HR1 with 1-day time bins.  For this time bin, the
typical relative uncertainty is 2\% for the count rate and 5\% for HR1.  The
PCA observations are indicated with  vertical lines.
\label{corr}}
\end{figure}

The correlation coefficient $r_{s1}$ between the X-ray count rate and
the hardness ratio HR1 for a 90 s time bin and 5.6-day correlation
intervals is shown in Fig.~\ref{corr}.  The results for different
correlation intervals are similar.  For longer intervals, the
correlations become statistically more significant as more data points 
are involved. The correlation interval of Fig.~\ref{corr} is the 5.6-day
orbital period of Cyg X-1. This allows us  to compare the result to the 
possible contribution from the orbital modulation of X-rays by the stellar
winds (\cite{wen99}). Both $r_{s1}$ and $r_{s2}$ (count rate vs HR2)
evolve similarly during the state transition.

The ASM count rate and spectral hardness in Fig.~\ref{corr} are positively
correlated in the soft state with an average $r_{s1}$ about +$0.7$,
corresponding to a false alarm possibility $<10^{-30}$. Significant
positive correlation was also found for time bins of 0.1 days, 1 day,
5 days and 10 days. In other words, the positive correlation in the soft
state holds for variability on time scales from 90 s up to at least 10
days. In the hard state, the correlation in Fig. ~\ref{corr} turns weak
and negative with an average $r_{s1}$ about $-0.2$.   Calculations
with time bins of 5.6 days (to eliminate the orbital effect), 20 days,
and 100 days also show negative correlation.  In both states, the
coefficient $r_{s2}$ behaves similarly but with relatively weaker
strength;  about 0.5 for the soft state and -0.12 for
the hard state for 90-s time bins and 5.6 d correlation intervals. 

The evolution of the correlation during the state transition is a
gradual one, as shown in Fig.~\ref{corr}. The entire episode of the
1996 state transition can be roughly divided into three distinct
phases: (1) a 15--20 day transition phase from the hard state to the soft
state, where $r_{s1}$ goes from negative to positive, (2) a $\sim$ 50 day 
soft state with a steady positive $r_s$, and (3) a 15--25 day transition 
phase back to the hard state. The start time of phase (1) and end time 
of phase (3) indicated in  Fig.~\ref{corr} were chosen to be the times 
when the hardness ratios are roughly at the mid-point between the mean 
levels of the hard and soft states. This yields a $\sim$ 20 day time 
scale for phases (1) and (3),  similar to what we would get if we 
were to choose the start (end) time to be when the correlation 
coefficient just started (ended) its sharp rise (drop). We obtain 
roughly the same results with smaller time bins (0.1 and 1 day), so
the conclusions seem quite robust.

\subsection{The PCA results}
The evolution of the two PCA hardness ratios with the energy flux is 
shown in Fig.~\ref{pca}.  As expected, the energy flux and spectral
hardness for $E < 12 $ keV (left panel) is strongly correlated in the
soft state (large filled circles) but weakly anti-correlated in the
hard state (small filled circles).  The evolution of the flux-hardness
correlation between the two states is apparent during  the transition
phases 1 and 3 (defined in Fig.~\ref{corr}) (open circles).  These
confirm  the ASM results.

To investigate possible orbital effects, we separated out the hard 
state data at phases 0.2--0.8, where phase 0 is defined as superior 
conjunction of the X-ray source. The results are shown in the inset
of Fig.~\ref{pca}. The anti-correlation seems to become less 
pronounced at low fluxes, indicating the importance of the orbital 
effects. On the other hand, the overall anti-correlation is still 
prominent, mostly due to the cluster of data points to the lower 
right, which come from observations 10-60 days proceeding or 
following the state transition (labeled with ``T''). Interestingly, 
one group of data points is from observations of Cyg X-1 during a 
50-day long soft flare that occurred about one year after the state
transition (labeled with ``F''). It is also worth noting that within the
remaining two data groups the correlation is absent (or very weak).

Data from the higher PCA energy bands ($E > 12 $ keV) show no strong
correlations in all states (right panel, Fig.~\ref{pca}).  There may 
be a slight anti-correlation in the hard state data, again mostly
due to observations near the state transition and during the soft 
flare. This is consistent with the fact that, within each state, the 
general shape of the observed spectrum above $\sim$12 keV seems insensitive to the 
change of the flux. 

\section{Discussion }
The ASM results show that the transition phases lasted for about 20
days during the 1996 state transition of Cyg X-1, as opposed to
$\lesssim$ 7 days as indicated by the change in the soft X-ray flux
(see Fig.~1).   That is, for about 20 days at the
beginning and near the end of the state transition episode, the system
was  in a transitional
process even though the soft flux was generally high. A similar conclusion was
drawn by Cui et. al (1997a, 1997b) based on the evolution of the
power density spectra (PDS).   We therefore suggest that the spectral states
of Cyg X-1 are better defined by the correlation between the soft
X-ray flux and the spectral hardness of the source than by the soft
flux alone.  

The PCA results confirm our findings based on the ASM data. 
Furthermore, they reveal that the orbital effects could account 
for the observed negative correlations on time scales less than 5.6 days, 
for time periods sufficiently far away from the soft flare or the state 
transition. This is consistent with the fact that
there is a broad absorption-like dip in the orbital light curves
obtained from the ASM data, likely caused by absorption and scattering
of the X-rays by the stellar wind (e.g., Wen et al. 1999).  Our
best-fit model ASM orbital light curve and hardness ratios (without
noise) would yield
$r_{s1}$ to be around -0.6 with a 0.1 day time bin in a 5.6 day
correlation interval. The observed correlation strength is weaker (
$r_{s1} \sim -0.2$), probably reduced by noise.  On the other hand,
the observed positive correlation in the soft state is the
opposite to what we would expect from the orbital effect.  This is
consistent with our previous conclusion (Wen et al. 1999) that orbital
modulation is much smaller (if present at all) in the soft state.

%  Indeed, correlation coefficients calculated indicate that
%  significant detections (<$10^{-4}$--$10^{-2} $ false alarm
%  possibilities) of the anti-correlation were mostly found at these phases.   

\begin{figure}[f]
\begin{centering}
\epsfxsize=6.5in \epsfbox{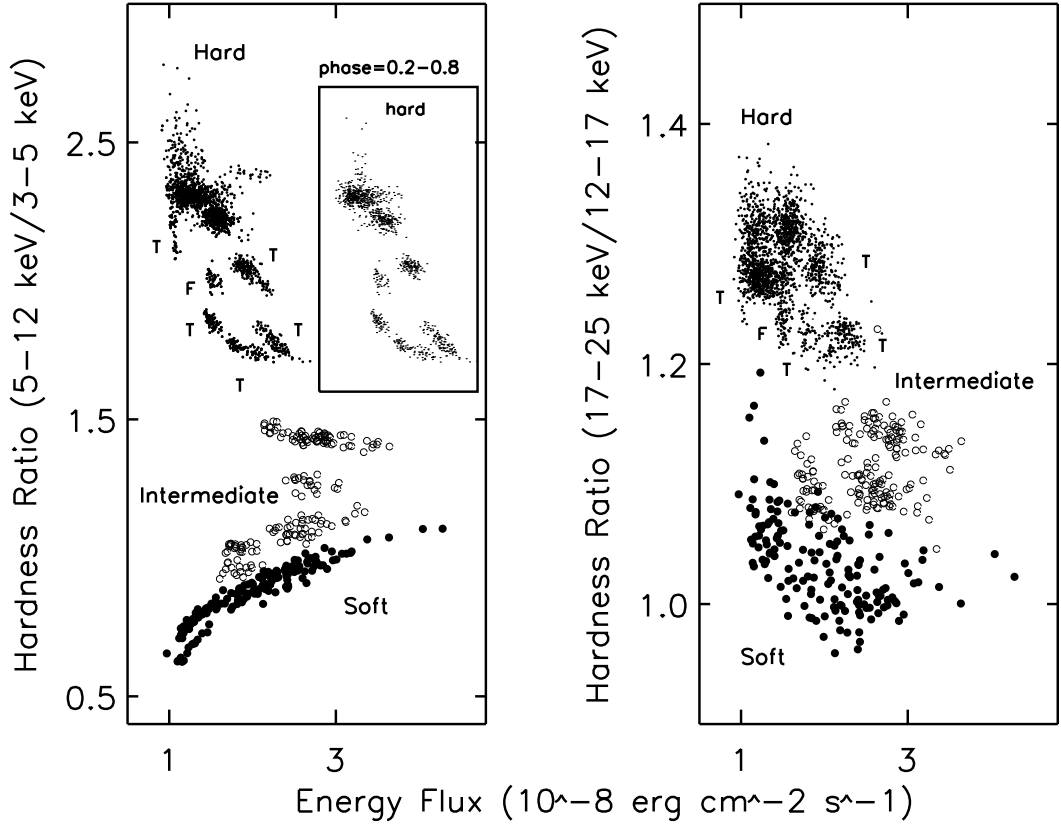}
\end{centering}
\figcaption[fig2.eps]{Evolution of the two PCA spectral hardness
ratios with the 2.5--25 keV incident energy flux (see text). 
Labels ``T'' and ``F''  indicate data  near the state transition and
 during the soft flare, respectively.  The
inserted window in the left panel shows the hard state data at phase
0.2--0.8 (with X-axis shifted to the right by one tick mark), where  phase 0 is superior conjunction of the X-ray
source. 
\label{pca}}
\end{figure}

Both the ASM and the PCA data indicate the existence of the
flux-hardness anti-correlation in the hard state on time scales longer
than the $5.6$ day orbital period.  Moreover, the hard state data near
the state transition episode and during the soft flare both contribute
to this observed anti-correlation in a similar fashion
(Fig.~\ref{pca}).  The PCA spectral fitting indicates that these
observations also share similar spectral properties, which deviate
somewhat from that of typical hard state data (see data section).
This seems to lend support to the notion that soft flares are ``failed''
state transitions. Similar soft flares occur randomly on time scales
of months to years in the ASM light curve. We therefore conclude that
the anti-correlation in the hard state on time scales longer than the
orbital period is intrinsic to the source and is probably related to
the mechanism that causes the state transition.

Models employing a simple disk--corona geometry predict that, as the 
mass accretion rate through the disk increases, the soft X-ray flux 
increases, which provides more seed photons for the 
inverse-Comptonization process and thus cools the hot electrons in 
the corona. This would be manifested observationally as the ``pivoting''
of the energy spectrum of the source. Such spectral pivoting is known
to occur in Cyg X-1 during a state transition (e.g., \cite{nolan84};
\cite{zhang97a}), and the pivoting energy is in the range of 10--20 
keV. This may explain the observed flux-hardness anti-correlation
at low energies (below the pivoting energy) for the state transition
(and perhaps soft flares). The same might also be true for the general
long-term evolution of the source in the hard state, if the pivoting 
phenomenon is universal (but more pronounced during a state transition). 
Interestingly, this may also explain the apparent lack of correlations 
around 12--25 keV (Fig. \ref{pca}), since the pivoting energy is right
in this energy range. Clearly, the same scenario cannot be applied to 
the soft state, where the correlation is observed to be (strongly) 
positive, unless the pivoting energy has moved to a very low energy 
(below the ASM pass band). Such a reduction in the pivoting energy 
is not apparent in our PCA data and is inconsistent with the lack of 
correlations in the 12--25 keV band. It is, therefore, likely that 
different physical processes are involved in the soft state 
(see \cite{zdziarski00} for a follow-up investigation of possible 
models).

% The pivoting behavior of the energy spectrum has been observed in
% the seyfert galaxy 3C 120 at a pivoting energy around 2--3 keV
% (\cite{maraschi91}), and for seed photon energies $\sim $ 10 eV.  

% Consistently, there has been observed in Cyg X-1 an anti-correlation
% of the powerlaw index and the flux in 45--140 keV range
% (\cite{crary96}) between data of high and low fluxes (presumally the
% hard and soft states).  

Li, Feng, \& Chen (1999) discovered similar correlations between the count
rate and hardness for Cyg X-1, using the PCA data, but on much shorter time
scales (0.01--100 s) and with little coverage in the hard state. In their study, they chose two energy bands:
2--6 keV and 13--60 keV, from which the hardness ratio was
derived. Such a choice of energy bands unfortunately masks the difference
in the correlations between the count rate and the spectral shape 
above and below 12 keV for the soft state. Using the BATSE data
(20-200 keV), on the other hand, Crary et al. (1996) did see a  lack
of correlation between the 45--140 keV energy flux and the photon index
within both data groups of high and low flux, which were presumed to
correspond to the hard and soft states, respectively. However, they
did not have the necessary soft-X-ray data to see the
count rate-hardness correlation that we found.  In this regards,  our
work bridges critical gaps in  those two
investigations and provides new insight into the overall spectral
behavior of this system. 

\acknowledgements

We acknowledge useful discussions with Andrzej Zdziarski and the
members of the ASM/{\it RXTE} team at MIT. This work was supported in part
by NASA through grants NAS5-30612 and NAG5-9098.

\pagebreak

\end{document}